\documentclass[aps,10pt,tightenlines,amsmath,onecolumn,amssymb]{revtex4}
\usepackage{graphicx}
\usepackage{dcolumn}
\usepackage{bm}

\begin{document}

\title{No axion-like particles from core-collapse supernovae?}

\author{Ignazio Bombaci}
\affiliation{Dipartimento di Fisica ``Enrico Fermi'', Universit\'a di Pisa and INFN, Sezione di Pisa, Largo Bruno Pontecorvo 3, 56127 Pisa, Italy}
\email{ignazio.bombaci@unipisa.it}

\author{Giorgio Galanti}
\affiliation{INAF, Osservatorio Astronomico di Brera, Via Emilio Bianchi 46, I -- 23807 Merate, Italy}
\email{gam.galanti@gmail.com}

\author{Marco Roncadelli}
\affiliation{INFN, Sezione di Pavia, Via A. Bassi 6, I -- 27100 Pavia, Italy, and INAF}
\email{marco.roncadelli@pv.infn.it}

\begin{abstract} 
A strong bound on the properties of axion-like particles (ALPs) has been set by assuming that ALPs are emitted by the protoneutron star just before the core-bounce in Galactic core-collapse supernovae, and that these ALPs subsequently convert to $\gamma$-ray photons which ought to be detected by a $\gamma$-ray mission. This argument has been applied to supernova 1987A to derive the bound on the ALP-photon coupling $g_{a \gamma \gamma} \lesssim 5.3 \cdot 10^{- 12} \, {\rm GeV}^{- 1}$ for an ALP mass $m_a \lesssim 4.4 \cdot 10^{- 10} \, {\rm eV}$, and can be applied to the next Galactic supernova to derive the even stronger bound $g_{a \gamma \gamma} \lesssim 2 \cdot 10^{- 13} \, {\rm GeV}^{- 1}$ for an ALP mass $m_a \lesssim 10^{- 9} \, {\rm eV}$. We carefully analyze the considered ALP production mechanism and find that it is oversimplified to an unacceptable extent. By taking into account the minimal ingredients required by a realistic analysis, we conclude that the previous results are doomed to failure. As a consequence, all papers quoting the above bound should be properly revised. Yet, since we are unable to rule out the possibility that protoneutron stars emit ALPs, in case a core-collapse supernova explodes in the Galaxy the $\gamma$-ray satellite missions active at that time should look for photons possibly coming from the supernova.
\end{abstract}


\maketitle


\noindent {\it Introduction} -- Axion-like particles (ALPs) are spin-zero, neutral and extremely light pseudo-scalar bosons (for a review, see~\cite{alp1,alp2}). As far as the present analysis is concerned, they are described by the Lagrangian 
\begin{equation}
\label{t1}
{\cal L}_{\rm ALP} = \frac{1}{2} \, \partial^{\mu} a \, \partial_{\mu} a - \, \frac{1}{2} \, m_a^2 \, a^2 + 
g_{a \gamma \gamma} \, a \, {\bf E} \cdot {\bf B}~,
\end{equation}
where $a$ is the ALP field, $m_a$ is the ALP mass while ${\bf E}$ and ${\bf B}$ denote the electric and magnetic components of the electromagnetic tensor $F^{\mu \nu}$. Manifestly, in the presence of an {\it external} electromagnetic field, the mass matrix of the photon-ALP system is off-diagonal, and so photon-ALP conversions $\gamma \to a$ take place~\cite{sikivie,anselm,rs1988}. The most up-to-date upper bound on the ALP-photon coupling $g_{a \gamma \gamma}$ is provided by the CAST (CERN Axion Solar Telescope) experiment and reads $g_{a \gamma \gamma} < 0.66 \cdot 10^{- 10} \, {\rm GeV}^{- 1}$ for $m_a < 0.02 \, {\rm eV}$ at the $2 \sigma$ level~\cite{cast}. Incidentally, exactly the same bound at the same confidence level has been obtained from the study of globular cluster stars~\cite{straniero}. 

Nowadays, ALPs have become popular mainly because of four different reasons. First, because they are a natural prediction of many extensions of the standard model, notably of superstrings and superbranes (see e.g.~\cite{turok1996,string1,string2,string3,string4,string5,axiverse,abk2010,cicoli2012,dias2014,scott2017} and references therein). Second, because for suitable values of $g_{a \gamma \gamma}$ and $m_a$ they are very good candidates for cold dark matter~\cite{preskill,abbott,dine,arias}. Third, because still for a range of values of $g_{a \gamma \gamma}$ and $m_a$ they would make the Universe considerably more transparent at very-high-energies (VHE, $100 \, {\rm GeV} < E < 100 \, {\rm TeV}$) than predicted by conventional physics, thereby allowing to probe the Universe deeper than expected~\cite{darma,bischeri,mm2009,prada2009,dgr2011}. Remarkably, the parameter space considered in the last point lends itself not only to astrophysical checks by the new generation of $\gamma$-ray observatories like CTA (Cherenkov Telescope Array)~\cite{cta}, HAWC (High-Altitude Water Cherenkov Observatory)~\cite{hawc}, GAMMA 400 (Gamma-Astronomy Multifunction Modules Apparatus)~\cite{g400}, LHAASO (Large High Altitude Air Shower Observatory)~\cite{lhaaso} and TAIGA-HiSCORE (Tunka Advanced Instrument for Gamma-ray and Cosmic ray Astrophysics-Hundred Square km Cosmic ORigin Explorer)~\cite{desy}. Fourth, because for the same values of $g_{a \gamma \gamma}$ and $m_a$ that would make the Universe more transparent ALPs can be detected in the near future with planned laboratory experiments like the upgrade of ALPS II~\cite{alps}, STAX~\cite{stax}, IAXO (International Axion Observatory)~\cite{iaxo} as well as with other devices~\cite{avignone1,avignone2,avignone3}.   

\begin{figure}[h]    
\centering
\includegraphics[width=0.40\textwidth]{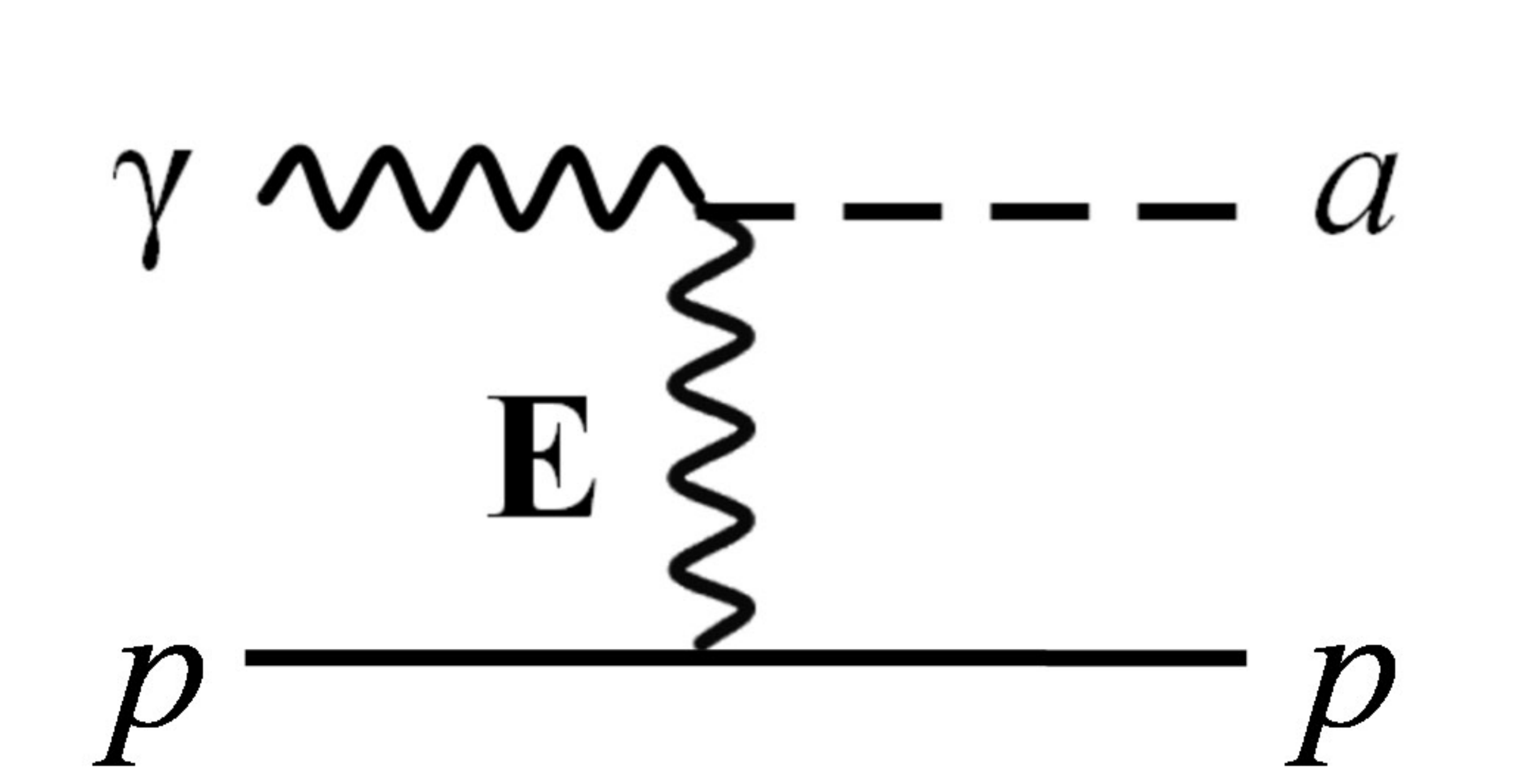}
\caption{\label{immagineA} Feynman diagram for the Primakoff process.}
\end{figure}

A hot topic concerning ALPs is their production in core-collapse supernovae. In 1996 two papers -- one by Brockway, Carlson and Raffelt (BCR)~\cite{raffelt1996} and another by Grifols, Mass\'o and Toldr\`a (GMT)~\cite{masso1996} -- appeared, which analyzed the ALP emission from supernova 1987A. Basically, the idea is as follows. The density of the protoneutron star just before and soon after the {\it core-bounce} is so high that a burst of ALPs is supposed to  be emitted through the Primakoff process $\gamma + p \to a + p$ -- sketched in FIG.~\ref{immagineA} -- almost simultaneously with the neutrinos which have been detected on Earth by the experiments IMB, Kamiokande II and IST~\cite{raffeltbook}.  

In the Galactic magnetic field ALPs should be converted into $\gamma$-ray photons of the same energy and ought to have been detected by the Gamma-Ray Spectrometer aboard the Solar Maximum Mission satellite (SMM), in spite of the fact that the line of sight of the SMM satellite was orthogonal to the direction of supernova 1987A. From the lack of detection of the corresponding photons a bound on the two-photon coupling 
$g_{a \gamma \gamma}$ has been derived. Subsequently, in 2015 Payez {\it et al.}~\cite{payez2015} published a very detailed follow-up paper on the same subject, reaching the conclusion that the upper limit on $g_{a \gamma \gamma}$ should be $g_{a \gamma \gamma} \lesssim 5.3 \cdot 10^{- 12} \, {\rm GeV}^{- 1}$ for an ALP mass $m_a \lesssim 4.4 \cdot 10^{- 10} \, {\rm eV}$, however without quoting any confidence level. Moreover, in 2017 Meyer {\it et al.}~\cite{mayer2017} published a Letter claiming that if a core-collapse supernova explodes in our galaxy then the {\it Fermi}/LAT detector can probe $g_{a \gamma \gamma}$ down to $g_{a \gamma \gamma} \simeq 2 \cdot 10^{- 13} \, {\rm GeV}^{- 1}$ for an ALP mass $m_a \lesssim 10^{- 9} \, {\rm eV}$. So, they conclude that `{\it it would allow an unprecedented exploration of the ALP parameter space for ALP masses below $10^{- 7} \, {\rm eV}$, surpassing current bounds and the projected sensitivity of future de\-di\-cated laboratory searches such as ALPS II and IAXO for masses up to $10^{- 8} \, {\rm eV}$}'. Finally, future $\gamma$-ray missions like e-ASTROGAM~\cite{astrogam}, AMEGO~\cite{amego}, ComPair~\cite{compair} and PANGU~\cite{pangu} could do even better owing to their higher sensitivity in the $(10 - 100) \, {\rm MeV}$ band with respect to {\it Fermi}. 

So, this means that the ALPs responsible for an enhanced transparency of the VHE Universe as well as for cold dark matter will either be detected or ruled out by relying upon their emission from a core-collapse supernova.

Given the importance of this issue it looks compelling to scrutinize all the assumptions upon which it rests. This is the aim of the present Letter.

\

\noindent {\it Previous work} -- In order to gain in clarity, we feel that the best approach is to start by going back to the pioneering work of Raffelt in which he considers the Primakoff process for ALP production by taking into account the Debye screening effects in an ordinary astrophysical plasma inside main-sequence and red giant stars~\cite{raffelt1986}. Basically, what happens is that a positively-charged ion $I_+$ attracts negatively-charged ions and repels positively-charged ones, so that outside a sphere of Debye radius $I_+$ exerts no electrostatic force. Accordingly, the Coulomb potential of $I_+$ is turned into a Yukawa-like potential. Moreover, Raffelt has computed the corresponding rate for ALP emission from main-sequence and red giant stars, which is given by 
\begin{equation} 
\Gamma_{\gamma \to a} = \frac{g^2_{a \gamma \gamma} \, T \, k^2}{32 \pi} \left[\left(1 + \frac{k^2}{4 \omega^2} \right) \, {\rm ln} \left(1 + \frac{4 \omega^2}{k^2} \right) \, - 1 \right]~,
\label{mr03122017a} 
\end{equation}
with $\omega$ the photon frequency -- incoming photons are assumed to have a blackbody spectrum -- and $k$ the Debye-H\"uckel screening scale in a non-degenerate plasma
\begin{equation} 
k = \frac{4 \pi \alpha}{T} \, \frac{\rho}{m_p} \left(Y_e + \sum_j Z_j \, Y_j  \right)~,
\label{mr04122017a} 
\end{equation}
where $T$ is the temperature, $\alpha$ is the fine structure constant, $\rho$ is the mass density, $m_p$ is the proton mass, $Y_e$ and $Y_j$ are the number fraction per baryon of electrons and heavy elements, respectively,  while $Z_j$ is the charge of the $j$-th element (this notation will be used also later and will be extended so that $Y_n$ denotes the number fraction of neutrons per baryon).  

Coming back to the case of the protoneutron star just before the core-bounce, the assumptions are very clearly stated in BCR. Electrons are completely degenerate. Hence, `{\it they are unavailable as scattering targets because their phase space is almost completely Pauli blocked}'. Moreover, `{\it they can be ignored for screening as well}'. Actually, $k$ is given by the Debye formula~\cite{raffelt1996}
\begin{equation} 
k = \frac{4 \pi \alpha}{T} \, n_p~,
\label{mr04122017b} 
\end{equation}
where $n_p$ is the number density of the protons: Eq. (\ref{mr04122017b}) is of course a particular case of Eq. (\ref{mr04122017a}). 

All three papers mentioned above assume that the photon mass is negligibly small, which is therefore set equal to zero. 

\

\noindent {\it Our critiques} -- As it is evident from the above short summary, the previous discussion is  actually framed within the vacuum apart from the condition that electrons are completely degenerate. Yet, a protoneutron star is a very complicated object (for a review, see~\cite{bombaciphysrep}). Below, we address some issues that should be taken into account in order to make a realistic description of ALP emission. 

We start by noting that it is apparent from Eq. (\ref{mr04122017b}) that protons are Debye screened by protons! But this fact is just the opposite of what happens in the usual Debye screening effect, and we do not understand what is going on, since protons repel each other (more about this, later). We would like to stress that this is a key-point of~\cite{raffelt1996,masso1996,payez2015,mayer2017}. 

Because the interest is focussed at a time just before the core-bounce, we observe that already for densities 
$\rho \simeq 2 \times 10^{12} \, {\rm g} \, {\rm cm}^{- 3}$ neutrinos are trapped in the stellar 
core~\cite{burrows-lattimer86,trapped} and weak interactions are in equilibrium ($\beta$-stable matter). Thus, to model in a realistic way the physical conditions of the stellar core at bounce, we consider 
$\beta$-stable nuclear matter with trapped neutrinos at constant entropy per baryon $S = 2$ (in unit of 
the Boltzmann constant) and constant total electron lepton number per baryon 
$Y_{Le} = 0.4$ \cite{burrows-lattimer86,bombaciphysrep}. In addition we include the effects of the nuclear interaction using the TM1-2 relativistic mean field model of~\cite{providencia2013}.

As an average core baryon number density we take $n_B = 0.32~{\rm fm}^{-3}$ (i.e. two times the value of the saturation density $n_B^0 = 0.16~{\rm fm}^{-3}$ of symmetric nuclear matter). For this value of the density we calculate the corresponding temperature and the particle number fractions, finding $T = 37.2 \, {\rm MeV}$ and $Y_n = 0.652$, $Y_p = Y_e = 0.348$, respectively. As a consequence, we are actually dealing with a globally neutral two-component plasma: electrons which are fully relativistic and fully degenerate, and protons and neutrons that are only slightly degenerate: for simplicity we ignore the latter fact and assume protons and neutrons to form a non-degenerate and 
non-relativistic gas. 

As we said, photons are assumed massless in previous works. However, even massless photons in a medium should be treated with a metric which takes into account the properties of the medium and differs from the usual Minkowski metric of ordinary vacuum (this point is explained in great detail in~\cite{raffeltbook}). 

Still, it is well known that in a (hot) plasma photons acquire a mass given by the plasma frequency, which in the present case reads 
\begin{equation} 
\omega_{\rm pl} = \bigl(\omega_{\rm pl, p}^2 + \omega_{\rm pl, e}^2 \bigr)^{1/2}~,
\label{mr04122017c} 
\end{equation}
where $\omega_{\rm pl, p}$ is the plasma frequency of the protons and $\omega_{\rm pl, e}$ is the plasma frequency of the electrons. Having assumed that protons are 
non-degenerate, we have~\cite{braatensegel}  
\begin{equation} 
\omega^2_{{\rm pl},p} = \frac{4 \pi \alpha \, n_p}{m_p} \left( 1 - \, 2.5 \, \frac{T}{m_p} \right) \simeq 75.32 \, {\rm MeV}^2~. 
\label{mr04122017d} 
\end{equation}
For electrons things are different. In spite of the fact that they are fully degenerate, they do contribute to the plasma frequency and according to Braaten and Segel~\cite{braatensegel} we have 
\begin{equation}
\label{b16052017a}
\omega^2_{{\rm pl},e} = \frac{4 \alpha}{3 \pi} \left(\mu^2 + \frac{\pi^2 T_c^2}{3} \right)~,
\end{equation}
where the chemical potential is
\begin{equation}
\label{b16052017b}
\mu = \left\{\left[\left(\frac{p_F^3}{2} \right)^2 + \left(\frac{\pi^2 T_c^2}{3} \right)^3 \right]^{1/2} + \frac{p_F^3}{2} \right\}^{1/3}  
\end{equation}
and $p_F = \left(3 \pi^2 n_e \right)^{1/3} \simeq 2.94 \cdot 10^2 \, {\rm MeV}$ is the Fermi momentum. A straightforward calculation yields 
\begin{equation}
\label{b16052017a}
\omega^2_{{\rm pl},e} \simeq 168.33 \, {\rm MeV}^2~,
\end{equation}
and so the plasma frequency follows by inserting Eqs. (\ref{mr04122017d}) and (\ref{b16052017a}) into Eq. (\ref{mr04122017c}), which gives $\omega_{\rm pl} \simeq 15.61 \, {\rm MeV}$. Thus, a photon -- or better to say a plasmon -- acquires a mass $m_{\rm pl} \simeq 15.61 \, {\rm MeV}$ which is just one-half of $T_c$, and so it is {\it by no means negligible} as compared to $T_c$, contrary to what has been assumed in previous works. Note that this fact sets a lower bound on the ALP emission cross-section. Incidentally, as explained in great detail in~\cite{raffeltbook}, massive photons in a plasma become longitudinal and transverse plasmons which obey different dispersion relations.   

Another point that has been totally ignored in the previous investigations is the presence of strong interactions. Because the whole protoneutron star is globally neutral and its average core density is larger than the nuclear saturation density, it can be regarded as a big, macroscopic nucleus (it is well known that ordinary nuclei cannot have a too large atomic number because of the Coulomb repulsion among protons). Now, this very fact has two implications. First, because in the present situation strong interactions are stronger than electromagnetic interactions, the concept of Debye screening loses any meaning. Second, it is an experimental fact that when a photon beam of energy larger than $10 \, {\rm MeV}$ but lower than the pion photo-production threshold strikes a rather heavy nucleus like Li, Be, C, O, AI, Si and Ca, a {\it giant resonance} occurs: photons are absorbed, excite collective modes of the nucleus which eventually breaks up~\cite{ahrens}. We expect a very similar situation to occur inside the protoneutron star. Photons should immediately thermalize, and so they have a blackbody spectrum at $T_c \simeq 37.2 \, {\rm MeV}$, which entails that their average energy is ${\overline E}_{\gamma} \simeq 100 \, {\rm MeV}$, hence just in the considered range. Therefore they ought to be absorbed by nuclear matter, thereby exciting collective modes: the consequence is not clear to us, even because after very few seconds the supernova explodes. Still, what is crucial is that -- getting absorbed -- photons are not anymore available for a possible conversion into ALPs. 

A further point of concern is that a magnetic field as large as $B \simeq \bigl(10^{10} - 10^{16} \bigr) \, {\rm G}$ in the stellar core is neglected altogether. A study of its consequences is beyond the scope of the present Letter. Nevertheless, two facts should be mentioned. One is that it will lead protons to emit synchrotron radiation. Because of the huge density, a very large self-absorption is expected. But some residual photons could survive. Depending on their energy, they might lift the degeneracy of the electrons in the highest Fermi levels, which would become free and therefore Debye screening the protons in the Primakoff process. The other fact is that such a very large magnetic field will presumably give rise to the same effect considered in~\cite{rs1988} in a somewhat different situation -- the photon-ALP conversion in the magnetosphere of a pulsar -- namely a suppression of the photon-ALP conversion.

So far, we have been dealing with nuclear matter as if it were in a normal state but it is totally unclear if 
it really makes sense to talk about protons and neutrons. Indeed, the transition from nuclear matter to quark-gluon plasma is known to occur at a temperature $T_{\rm trans} \simeq 200 \, {\rm MeV}$, but in the protoneutron star core there is a pressure of $10^{35} \, {\rm dyne} \, {\rm cm}^{- 2}$ which presumably favors such a transition, which accordingly could take place at a smaller temperature. As a consequence, one might be led to replace neutrons and protons by quarks and gluons. This is however still an open question even for an even denser neutron star~\cite{qgp1,qgp2,qgp3}.

A final remark is in order. In Meyer {\it et al.}~\cite{mayer2017} the following detection strategy is envisaged: `{\it The ALP-induced $\gamma$-ray signal is expected to arrive roughly simultaneously to the neutrinos and hence the neutrinos signal would provide the required timing information to search for a coincident $\gamma$-ray signal}'. We simply disagree. The reason is as follows. ALPs -- if produced -- they immediately escape. To see this, the only possible interactions of ALPs (denoted by $a$) with radiation or matter (denoted by $f$) are $a + \gamma \to a + \gamma$, $a + \gamma \to f + \overline{f}$ and $a + f \to \gamma + f$. Now, taking into account the CAST bound, a simple estimate of these processes yields $\sigma (a + \gamma \to a + \gamma) < \left({\cal E}^2/{\rm GeV}^2 \right) 10 ^{- 68} \, {\rm cm}^2$, $\sigma (a + \gamma \to f + {\overline f}) \sim \sigma (a + f \to \gamma + f) < 10^{- 50} \, {\rm cm}^2$ (${\cal E}$ is the ALP energy), from which we see that -- since the maximal photon energy is $100 \ {\rm MeV}$ -- the first process is negligible. Even assuming that both the temperature and the baryon density are uniform for the mean free paths we get $\lambda (a + f \to \gamma + f) > 10^{11} \, {\rm cm}$ and $\lambda (a + \gamma \to f + {\overline f}) > 10^{14} \, {\rm cm}$ (this situation is completely different from the one envisaged in~\cite{turner}, since in that case axion trapping -- and not ALP trapping -- is due to Yukawa coupling of axions). On the other hand, neutrinos are trapped during the phase of protoneutron star until about 0.5 s after the supernova explodes. Thus, we do not see any time correlation between ALP emission -- and so the ALP induced $\gamma$-ray signal -- and the bulk of neutrino emission.

\

\noindent {\it Conclusions} -- We have critically considered the previous treatments of ALP emission from a protoneutron star just before the supernova explosion. We have shown that those treatments are unacceptably oversimplified, and that at the same time too many aspects of the problem are still not well understood to such an extent that a proper treatment seems presently beyond any hope. What is nevertheless clear is that all previous results are doomed to failure.

As a consequence, all papers quoting the bound derived by Payez {\it et al.}~\cite{payez2015} should be properly revised.

Quite recently, a paper concerning the possible coupling of ALPs to electrons has appeared, in which an upper bound has been derived from supernova 1987a~\cite{chang2018}. This is totally irrelevant for our considerations, since it concerns a mass range $(1 - 10^3) \, {\rm MeV}$.

Yet, since we are not in position to rule out the possibility that protoneutron stars emit ALPs, it is a very good thing that in case a core-collapse supernova explodes in the Galaxy the 
$\gamma$-ray satellite missions active at that time look for photons possibly coming from the supernova.

\

\noindent {\it Acknowledgments} -- M. R. thanks Maurizio Giannotti, Ettore Fiorini, Carlotta Giusti, Georg Raffelt, Oscar Straniero and in particular Roberto Turolla for discussions about topics concerning the present Letter and Giancarlo Setti for reading and commenting a preliminary draft. The work of I. B. is supported by an INFN NEUMATT grant, G. G. acknowledges contribution from the grant INAF 
CTA--SKA, `Probing particle acceleration and $\gamma$-ray propagation with CTA and its precursors', while the work of M. R. is supported by an INFN TAsP grant.

\end{document}